\documentclass{aa}

\usepackage{enumitem}
\usepackage{graphicx}
\usepackage{txfonts}
\usepackage{natbib}
\bibpunct{(}{)}{;}{a}{}{,}

\usepackage{lscape}
\usepackage{longtable}
\usepackage{amssymb}

\usepackage{hyperref}

\usepackage{color}

\begin{document}
 
 \title{Gas infall via accretion disk feeding Cepheus\,A HW2}

  \author{A. Sanna\inst{1,2}, A. Oliva\inst{3,4}, L. Moscadelli\inst{5}, C. Carrasco-Gonz\'{a}lez\inst{6}, A. Giannetti\inst{7}, G. Sabatini\inst{5}, M. Beltr\'{a}n\inst{5}, 
  C. Brogan\inst{8}, T. Hunter\inst{8}, J.\,M. Torrelles\inst{9,10}, A. Rodr\'{i}guez-Kamenetzky\inst{11}, A. Caratti~o~Garatti\inst{12}, R. Kuiper\inst{13}}


                                      

   \institute{INAF, Osservatorio Astronomico di Cagliari, via della Scienza 5, 09047, Selargius, Italy \\
   \email{alberto.sanna@inaf.it}
   \and Max-Planck-Institut f\"{u}r Radioastronomie, Auf dem H\"{u}gel 69, 53121 Bonn, Germany
   \and D\'{e}partement d'Astronomie, Universit\'{e} de Gen\`{e}ve, Chemin Pegasi 51, CH-1290 Versoix, Switzerland
   \and Space Research Center (CINESPA), School of Physics, University of Costa Rica, 11501 San Jos\'{e}, Costa Rica
   \and INAF-Osservatorio Astrofisico di Arcetri, Largo E. Fermi 5, 50125 Firenze, Italy
   \and Instituto de Radioastronom\'{i}a y Astrof\'{i}sica UNAM, Apartado Postal 3-72 (Xangari), 58089 Morelia, Michoac\'{a}n, M\'{e}xico
   \and INAF - Istituto di Radioastronomia \& Italian ALMA Regional Centre, Via P. Gobetti 101, I-40129 Bologna, Italy
   \and NRAO, 520 Edgemont Road, Charlottesville, VA 22903, USA
  \and Institut de Ci\'{e}ncies de l'Espai (ICE-CSIC), Can Magrans s/n, 08193, Bellaterra, Barcelona, Spain
  \and Institut d'Estudis Espacials de Catalunya (IEEC), Barcelona, Spain
  \and Instituto de Astronom\'{i}a T\'{e}orica y Experimental (IATE, CONICET-UNC), C\'{o}rdoba, Argentina
  \and INAF - Osservatorio Astronomico di Capodimonte, salita Moiariello 16, 80131, Napoli, Italy
  \and Faculty of Physics, University of Duisburg-Essen, Lotharstra$\rm \beta$e 1, D-47057 Duisburg, Germany
}
   \date{Received ...; accepted ...}


  \abstract{The star-forming region Cepheus\,A hosts a very young star, called HW2, that is the second closest to us growing a dozen times more massive than
  our Sun. The circumstellar environment surrounding HW2 has long been the subject of much debate about the presence or not of an accretion disk,  whose
  existence is at  the basis of our current paradigm of star formation. Here, we answer this long standing question by resolving the gaseous disk component,
  and its kinematics, through sensitive observations at cm wavelengths of hot ammonia (NH$_3$) with the Jansky Very Large Array. We map the accretion
  disk surrounding HW2 at radii between 200 and 700\,au, showing how fast circumstellar gas collapses and slowly orbits to pile up near the young star at very
  high rates of $2\times10^{-3}$\,M$_{\odot}$\,yr$^{-1}$. These results, corroborated by state-of-the-art simulations, show that an accretion disk is still efficient in
  focusing huge mass infall rates near the young star, even when this star has already reached a large mass of 16\,M$_{\odot}$.}
  

   
   
   


   \authorrunning{A. Sanna et al.}
   \maketitle
%

\section{Introduction}

The large reservoir of interstellar gas needed to build up a massive star, dozens of times more massive than our Sun, piles up over  wide regions of the order of
a parsec (pc). But, it is only within circumstellar regions as large as a few hundred times an astronomical unit (au) that gas will be ultimately collected to accrete
onto a ``small'' proto-star, with diameter of about a million kilometers only. Resolving the properties of the gas flow, as it streams in the inner hundreds au from a
very young star, has long been an observational challenge, especially for the most massive stars which are found much further away from Earth than Solar-type
stars \citep[e.g.,][]{Cesaroni2007,Beltran2016}. The circumstellar regions where gas collapses onto a plane, and orbits faster as it is pulled inward by the central
star's gravity, are generally referred to as the accretion disk, which is the physical structure conveying mass onto Solar-type stars ultimately  \citep[e.g.,][]{Dauphas2011}.
Since decades, this general picture of disk-mediated accretion has been much questioned when applied to the formation of massive stars, where a large 
mass reservoir of several tens of Solar masses has to focus near the star and sustain mass infall rates exceeding 10$^{-4}$\,M$_{\odot}$\,yr$^{-1}$. Under these
conditions, disk stability can be severely affected by, for instance, local fragmentation, tidal interactions with nearby cluster members and powerful stellar feedback,
as suggested both theoretically and observationally \citep[e.g.,][]{Meyer2018,Oliva2020,Maud2017,Goddi2020,Johnston2020,Sanna2021}. As such,  not even for
those massive stars at smaller distances from the Sun we have reached a general consensus yet.

At a trigonometric distance of 700\,pc from the Sun \citep{Moscadelli2009,Dzib2011}, Cepheus\,A is the second nearest star-forming site where massive young stars
of ten and more Solar masses  are born \citep{Kun2008}. In the innermost regions of Cepheus\,A, an outstanding bright and compact radio source, called ``HW2'' after
Hughes \&  Wouterloot \citep{Hughes1984,Garay1996,Rodriguez1994,Curiel2006}, was soon recognized to pinpoint a newly born massive star, whose own
luminosity dominates that of the whole region of several thousands Solar luminosities \citep[e.g.,][]{DeBuizer2017}. The environment surrounding HW2 has been the target
of many studies aimed to verify star formation theories in past years, because its close distance to  the Sun provides a preferred test laboratory. Nevertheless, these
studies have been contradictory at the very least, once suggesting the presence of a circumstellar disk \citep[e.g.,][]{Patel2005,Torrelles2007,JimenezSerra2009,Ahmadi2023} 
and the other questioning it as due to a chance superposition of different protostellar cores \citep[e.g.,][]{Brogan2007,Comito2007}. At the bottom line, two main questions
were left open, whether HW2 forms via disk accretion or, alternatively, whether this lack challenges our understanding of how massive stars gather their large masses when
still very young \citep[e.g.,][]{Goddi2020}.

In late spring 2019, we performed high sensitivity, Very Large Array (VLA) observations at wavelengths near 1\,cm of the hot ammonia (NH$_3$) line emission,
that is excited in the circumstellar gas around HW2. When talking about hot gaseous ammonia, one refers to emission from those transitions excited at kinetic
temperatures above 100\,K, such as the metastable  (J\,$=$\,K) inversion transitions with J higher than 3 observable at radio wavelengths \citep[e.g.,][]{Ho1983}.
Their detection is a signpost of dense and warm gas heated in the vicinity of a young massive star \citep[e.g.,][]{Cesaroni1992,Goddi2015}. The idea behind this
project was to search via hot ammonia for direct evidence of an accretion disk surrounding HW2, by resolving physical scales of the order of 100\,au.


\section{Observations \& data reduction}

We conducted spectroscopic VLA observations in band K (18.0--26.5\,GHz) of the Galactic star-forming region Cepheus\,A. Observations were carried out 
under program 19A-248 on 2019 at four epochs of 2.5 hours each, on April 27, June 24, 26, and 27. The array observed with 27 antennas in B configuration
during the first three epochs, with baselines ranging from a minimum of 10\,k$\lambda$ to a maximum of 830\,k$\lambda$ approximately. At the last epoch,
the array started moving to the largest configuration with baselines extending up to 1.5\,M$\lambda$. Observations information is summarised in
Table\,\ref{tobs}.

We made use of a mixed correlator (WIDAR) setup consisting of 3 basebands, a single baseband sampled at 8 bits and 1.0\,GHz wide, and two basebands
sampled at 3 bits of 2.0\,GHz each  \citep{Perley2011}. These basebands were centered at frequencies of 24.170, 22.210, and 20.000\,GHz respectively. The 
first setup was used to cover the NH$_3$\,(1,\,1) to (5,\,5) metastable inversion transitions (Table\,\ref{tlines}) with five spectral windows, each 16\,MHz wide,
and 384 channels of 41.7\,kHz. These settings provide a velocity coverage of 200\,km\,s$^{-1}$ near each line, wide enough to encompass their hyperfine
structure, and a narrow velocity resolution of 0.5\,km\,s$^{-1}$ to sample the range of ammonia emission, of the order of 10\,km\,s$^{-1}$ \citep{Torrelles2007}.
Spectral windows of 128\,MHz and 64 channels were used to fill all basebands for continuum measurement.

The VLA visibility data were calibrated with the Common Astronomy Software Applications (CASA) package, version\,5.3.0 \citep{CASA2022}, and processed through
the scripted  pipeline, version\,1.4.2, which uses the Perley-Butler 2017 flux density scale. Manual flagging was performed on each scheduling block separately. We
imaged the five spectral line windows with the task \emph{tclean} of CASA combining the four scheduling blocks and processing the $uv$ data with natural weightings,
to maximize the sensitivity to the ammonia emission. Image cubes were used to select the line-free channels from a spectrum integrated over a circular area of
$3''$ in size, which was centered on the target source. These channels were input into the task \emph{imcontsub} of CASA to subtract a constant continuum
level (\emph{fitorder}\,$=$\,0) from the image cubes, obtaining a continuum-free dataset for each ammonia line. All spectral line images discussed in the paper were
obtained by applying natural weightings, to maximize the array sensitivity and retrieve faint emission at increasing distances from the star. The image produced with
continuum spectral windows was self calibrated to improve on the map dynamic range (peak of 11\,mJy\,beam$^{-1}$) and uses Briggs's robust weightings ($\Re$\,$=$\,1)
to minimize beam side lobes. 

Synthesized beams are $0\farcs452$\,$\times$\,$0\farcs285$ at $-75.8\degr$ and $0\farcs512$\,$\times$\,$0\farcs306$ at $-71.9\degr$ for the line 
and continuum maps, respectively, with position angles measured east of north. Line imaging achieves a rms thermal noise of approximately 0.9\,mJy\,beam$^{-1}$
per resolution unit, corresponding to a brightness temperature of about 15\,K at the given beam. Continuum imaging achieves a rms thermal noise of approximately
5\,$\mu$Jy\,beam$^{-1}$ with the continuum central frequency set at 21.8\,GHz.

\begin{figure}
\centering
\includegraphics [angle= 0, scale= 0.8]{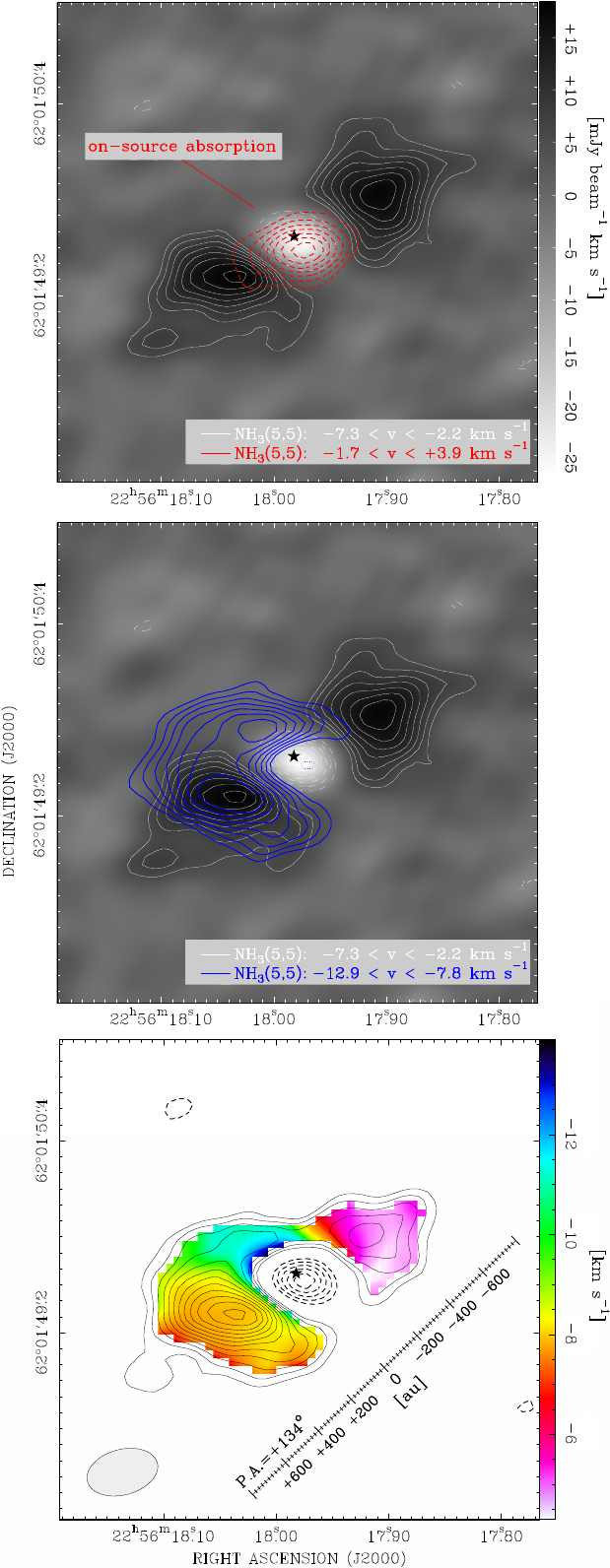}
\caption{Accretion disk in ammonia surrounding HW2 (black star). \textbf{Top:} moment zero maps of the (5, 5) inversion line evaluated within two velocity ranges,
as indicated on the bottom right  (beam size on the bottom panel). White contours and grey background quantify the line brightness about the systemic velocity,
whereas the red dashed contours the red-shifted absorption observed against the central region. Contours start at 3\,$\sigma$ increasing by 1\,$\sigma$ of 
1.8\,mJy\,beam$^{-1}$\,km\,s$^{-1}$; dashed (negative) contours start at $-3\,\sigma$ decreasing by $2\,\sigma$. \textbf{Middle:} similar to the top panel, but with
the moment zero map at red-shifted velocities replaced by one at blue-shifted velocities. \textbf{Bottom:} moment zero (black contours) and first moment (colors)
maps obtained by combining the systemic and blue-shifted ranges, with velocities quantified by the color wedge to the right. Same contour levels as previous panels,
with 1\,$\sigma$ of 2.6\,mJy\,beam$^{-1}$\,km\,s$^{-1}$. A linear scale is drawn along a position angle of 134$\degr$ east of north and all panels show the same 
field.} \label{fig1}
\end{figure}

\section{Results}

In Fig.\,\ref{fig1}, we present three plots summarizing the kinematic information inferred from the ammonia line imaging.  Firstly, we identified three ranges of velocity,
 5 to 6\,km\,s$^{-1}$ wide, that are associated with three, different, spatial distributions of gas near the young star. These velocities will be referred to as the redshifted
(from $+3.9$ to $-1.7$\,km\,s$^{-1}$), the systemic (from $-2.24$ to $-7.3$\,km\,s$^{-1}$),  and the blueshifted ranges (from $-7.8$ to $-12.9$\,km\,s$^{-1}$).  All
together, they encompass the whole range of emission of dense molecular gas, from $+4$ to $-13$\,km\,s$^{-1}$ approximately, as reported in past, larger scales
observations \citep[e.g.,][and references therein]{JimenezSerra2009}.

Beginning with the upper panel, we compare two maps of the NH$_3$\,(5, 5) emission integrated in velocity (namely, a moment-zero map), that are obtained from
two distinct velocity intervals taken consecutive (marked in red and white). The interval at higher velocities (red dashed contours) is characterized by molecules
observed in absorption only, whose peak is centered on top of the previously identified star position \citet{Curiel2006}. These velocities are redshifted with respect to
the systemic velocity (V$_{sys}$) of the region  at nearly $-4.5$\,km\,s$^{-1}$ \citep{Sanna2017}. In the upper and middle panels of Fig.\,\ref{fig1}, white contours
are used to draw emission integrated over the velocity range centered on the systemic velocity and 6\,km\,s$^{-1}$ wide. This emission is characterized by two
approximately symmetric lobes (hereafter, the bulk emission) to the southeast and northwest of the central absorption region, which is still visible at these velocities
(dashed white contours). In the redshifted map, the absorption region has an elongated spatial morphology connecting the bulk emission to the south of the star
position, with a deconvolved (Gaussian) size of $0\farcs26$\,$\times$\,$0\farcs06$  at a position angle of $101\degr\pm4\degr$. In the middle panel, the integrated
map at redshifted velocities is replaced by one at blueshifted (lower) velocities where ammonia is observed in emission only (blue contours). The blueshifted emission
is characterized by an arched spatial morphology that bridges the bulk emission to the north of the star position.

\begin{figure*}
\centering
\includegraphics [angle= 0, scale= 1.2]{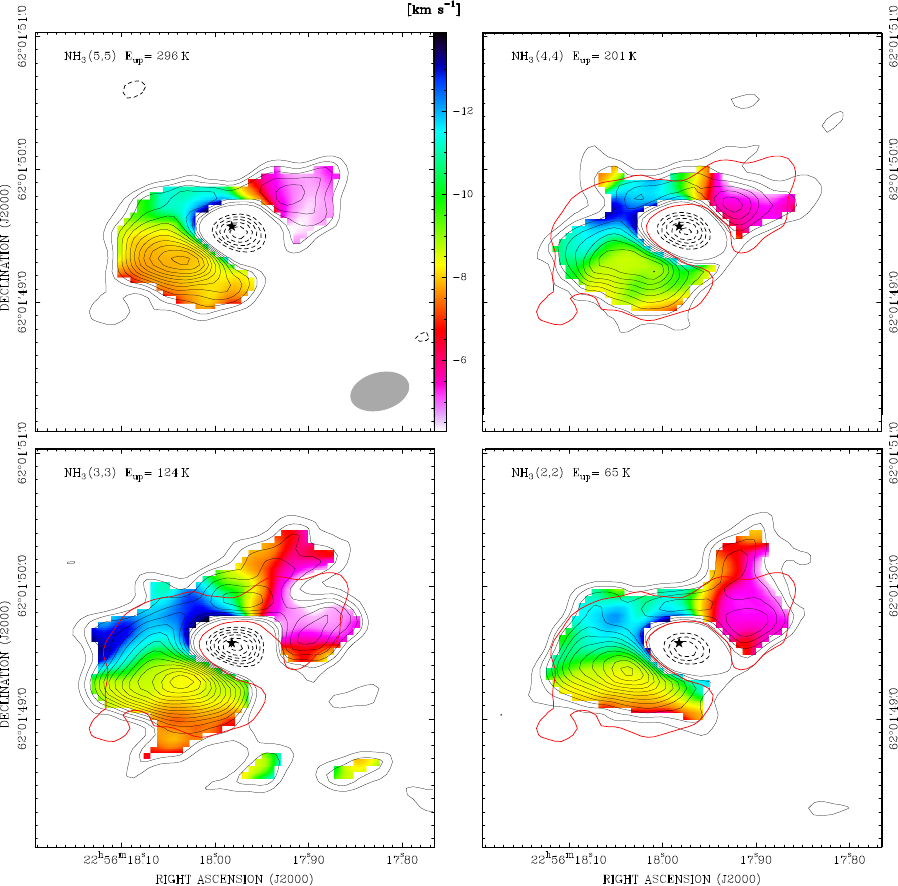}
\caption{Comparison of the NH$_3$ gas kinematic among the inversion lines from (2, 2) to (5, 5), as indicated in the top left with upper 
excitation energies. Each panel shows a first moment map of a given line (color map) with overplotted its moment zero contours.  Moments
are calculated over the same velocity range (see Fig.\,\ref{fig1}) and the velocity scale is quantified by the color wedge in the top left panel.
Positive contours start at 3\,$\sigma$ and increase by 1\,$\sigma$ of 2.6\,mJy\,beam$^{-1}$\,km\,s$^{-1}$; negative (dashed) contours start at
$-3\,\sigma$ by steps of $-2\,\sigma$. First moment maps are evaluated within the 4\,$\sigma$ contour of their moment zero map; the red thick
contour draws the 3\,$\sigma$ contour of the NH$_3$\,(5, 5) map, for comparison with other lines. Same field of view and resolution as in
Fig.\,\ref{fig1} (beam size on the top left panel).} 
\label{fig2}
\end{figure*}

The bottom panel delivers full kinematic information of the bulk and blueshifted velocities integrated together, through a (colored) first-moment map overplotted
on moment-zero (black) contours. The moment-zero map shows a nearly complete ring of ammonia emission surrounding the absorption region: moving outward
from the absorption center, the ring has a minimum radius of nearly 200\,au and can be detected up to radii between 600 and 700\,au, at the current sensitivity (cf.
linear scale in the panel). The first-moment map is evaluated within the 4\,$\sigma$ contour of the moment-zero map and its local weighted velocity is quantified by
the color scale to the right, ranging from $-13.9$ to $-4.3$\,km\,s$^{-1}$. There is a regular change of gas velocity across the ring whose apparent pattern
is the combination of two velocity components, counterclockwise rotation and infall motions, as it will be demonstrated later on. We can show
that a similar velocity pattern is observed with different ammonia lines at decreasing excitation energies, as proof of an intrinsic gas property (Fig.\,\ref{fig2}).
Ammonia emission appears fainter to the southwest in correspondence of the redshifted absorption, where it is only faintly detected through the (2,\,2) and (4,\,4)
inversion lines showing a closed ring. In the following, we will generally refer to the NH$_3$ molecular ring as the accretion disk.

One can demonstrate that the accretion disk encircles the region near HW2 where the bright radio emission was discovered, by overplotting maps of ammonia and
centimeter continuum (Fig.\,\ref{figB1}). The latter is a tracer of thermal bremsstrahlung emission from hydrogen gas ionized along a protostellar jet. This radio
thermal jet, a prototypical example of the kind firstly resolved in the early '90 \citep{Rodriguez1994}, was recently shown to arise at physical scales of only 20\,au
from HW2 \citep{Carrasco2021}. The correspondence between the molecular ring center with the bright radio continuum from the protostellar jet provides direct 
explanation for the ammonia absorption, which arises from gas distributed further away from the star and in foreground against the continuum. This feature, as well
as those discussed in the following, are explained in the cartoon of Fig.\,\ref{fig3}.

\subsection{NH$_3$ line fitting}

In order to derive the physical conditions of ammonia gas, we made use of the brightness ratio between three inversion transitions, the (2,\,2), (4,\,4),
and (5,\,5). Their state belongs to the subgroup of NH$_3$ molecules whose hydrogen spins are antiparallel, that are named para-NH$_3$ and correspond
to quantum numbers $K\ne3n$ ($n$ an integer). Alternatively, states with all hydrogen spins parallel are named ortho-NH$_3$   ($K=3n$). Changes 
of spin orientation are not allowed via radiative or collisional transitions, which makes of para- and ortho-NH$_3$ two distinct species that do not 
mix \citep[e.g.,][]{Ho1983}. For this reason, the (3,\,3) inversion transition was neglected from the spectral fitting. For our purposes, emission coming 
from the (1,\,1) inversion transition was also excluded due to its low excitation energy, of only 23.8\,K, which implies heavy contamination from cold
gas belonging to the diffuse envelope at large distances from the star \citep[e.g.,][]{Mangum2013}.

We extracted three spectra per transition from three distinct circular areas with diameter equal to the beam major axis (of 316\,au), which optimizes
signal-to-noise ratio and angular resolution (Fig.\,\ref{fig4}). We drew the best fit to the line passing through the absorption peak and the peaks of the
bulk emission and centered the pointings along this line. In order to compare the lobes of the bulk emission, two pointings were equally spaced by
380\,au from the absorption peak, where a third pointing was set (grey shades in Fig.\,\ref{fig4}). Spectral resolution was Hanning smoothed to improve
on the sensitivity.

In Figs.\,\ref{figA1} and \ref{figA2}, we plot four spectra including the (3,\,3) inversion transition, to emphasize the existence of two, overlapping, velocity components
at the position of absorption, that are detected both in the para and ortho species  (single components detected elsewhere). Each spectrum shows similar
properties consisting of two blended peaks of absorption, between $-8$ and $+4$\,km\,s$^{-1}$, with a first component peaking near the negative
systemic velocity (V$_{sys}$) and a second component at positive redshifted velocities. These two components will be treated separately in the following.
Notably, in Fig.\,\ref{figA1} we also plot theoretical positions of the satellite hyperfine lines of NH$_3$ for each velocity component (green and magenta
separately), as computed from analytic formulas in \citet{Kisliuk1950}. Hyperfines are neglected in the following because heavily affected by the noise,
which prevents their positive detection and fit.

We applied the same approach for spectral fitting followed by \citet{Sanna2021} which also accounts for the continuum level. Firstly, we searched for best 
guesses of the gas physical conditions by  means of Weeds, a package of the Grenoble Image and Line Data Analysis Software (GILDAS $-$ \citealt{Maret2011}),
trying to reproduce the NH$_3$ spectra both in emission and absorption.  Weeds generates synthetic spectra of a given molecular species
whose energy levels are populated by thermal distribution, after fixing five parameters: \textbf{(i)} the intrinsic full width at half maximum (FWHM) of the
spectral lines;  \textbf{(ii)} the line offset with respect to the rest velocity in the region; \textbf{(iii)} the spatial extent (FWHM) of the emitting region,
approximated to a Gaussian brightness profile; \textbf{(iv)} the rotational temperature and \textbf{(v)} column density of the emitting gas. Parameters
\textbf{(i)} and \textbf{(ii)} were tested separately by fitting a Gaussian profile to each inversion line.

Secondly, we processed the initial parameters into MCWeeds exploiting a Bayesian statistical analysis for an automated fit of the spectral lines
\citep{Giannetti2017}. We minimized the difference between the observed and computed spectra  via Monte Carlo Markov chains deriving best
fit parameters with their statistical uncertainties. To reproduce the central absorption, we set the background continuum temperature based on the
22.2\,GHz flux density inside the pointing area (of 7.3\,mJy), but assign an effective size equal to the A-array beam (of 80\,mas). In doing so, we
account that 22.2\,GHz observations by \citet{Curiel2006}  detected same peak fluxes (within uncertainties) at the resolution of the longest VLA
baselines, as evidence of compact emission. This assumption is further supported by the fit convergence a posteriori.

The following (standard) caveats also apply. Synthetic spectra are produced under the working hypothesis of a single excitation temperature and
column density (per velocity  component), and are assumed to be uniform along the line-of-sight that, on the contrary, crosses portions of gas at different
distances from the star, where one expects a smooth change of the gas physical conditions (e.g., streamline in Fig.\,\ref{fig3}). Different transitions
are also forced to have same linewidth, assuming that each spectral component is excited within the same parcel of gas. These approximations will
be disregarded in our analysis due to the negligible discrepancies between observed and fitted line profiles.

In Table\,\ref{tpar}, we list the results of the NH$_3$ spectral fitting for the three pointings to the eastern side, western side, and in front of the
young star (column~1);  columns~2 and~3 quantify the (projected) linear distance of each pointing from the star position and extent of the
integrated areas, respectively; columns~4 to~7 list the best fit parameters of the gas emission output by MCWeeds with their statistical uncertainties;
column~8 specifies the opacity computed by Weeds for the NH$_3$ (5,\,5) line imaged in the main figures. In Fig.\,\ref{figA2}, we also show the 
same absorption spectra of Fig.\,\ref{figA1} but including the continuum level for completeness.

\section{Discussion}

\begin{figure}
\centering
\includegraphics [angle=0, scale= 0.25]{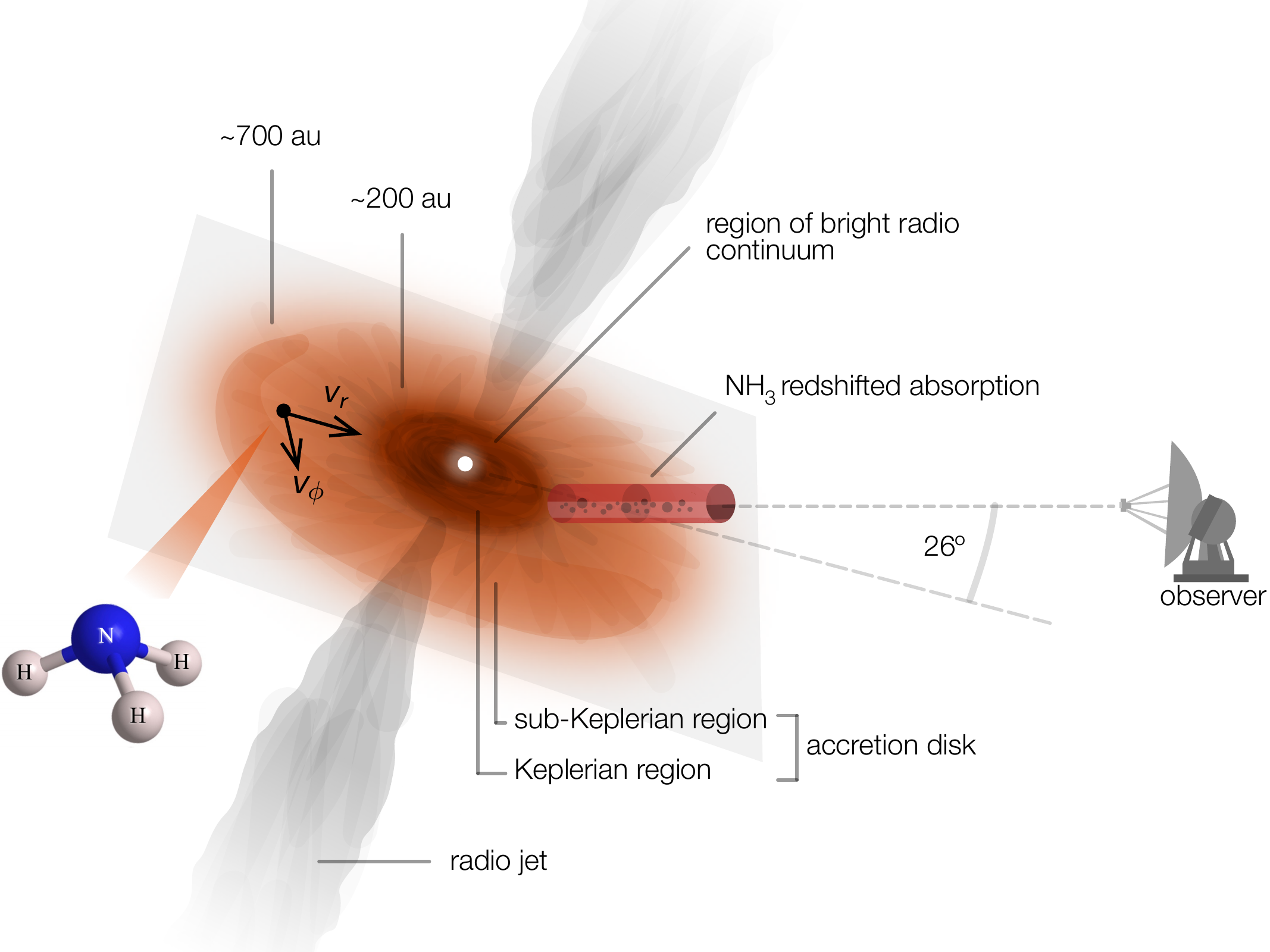}
\caption{Cartoon of the star formation scenario around HW2, where the observer is placed to the right for clarity. The line-of-sight to HW2, slightly above the disk 
mid-plane by $26\degr$ \citep{Sanna2017}, crosses a column of NH$_3$ gas infalling in front of a region of bright continuum emission, confined inside the inner
200\,au. At larger radii, the observer maps emission from NH$_3$ gas  (molecular structure sketched to the left) which is both infalling (v$_{\rm r}$) by gravity
toward the central star and rotating (v$_{\phi}$) at sub-Keplerian velocities. Atomic (ionized) gas is also ejected away along the general direction of the system
angular momentum, with a lobe of blueshifted (approaching to the observer) gas to the north and one of redshifted (receding) gas to the south \citep[e.g.,][]{Gomez1999}.}
\label{fig3}
\end{figure}

\begin{figure*}
\centering
\includegraphics [angle=0, scale= 0.85]{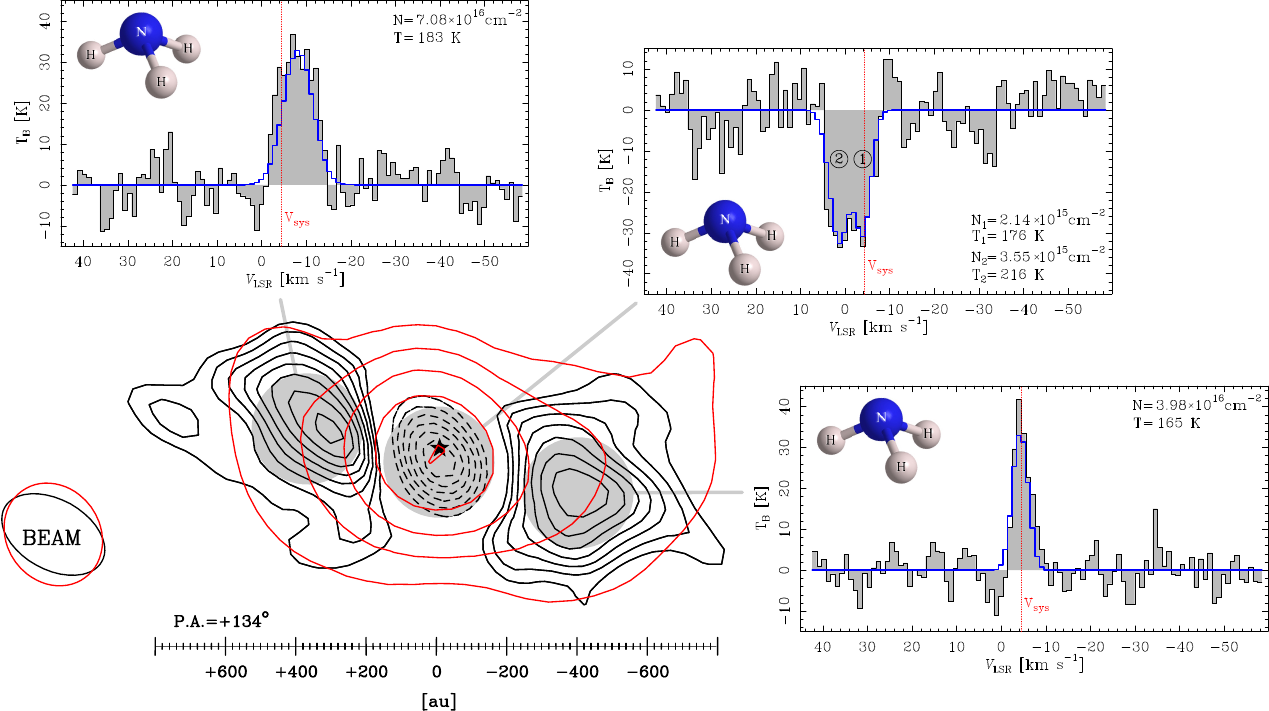}
\caption{Gas physical conditions near HW2. \textbf{Bottom left:} Spatial extent of the NH$_3$\,(5,\,5) bulk emission (black contours, same as Fig.\,\ref{fig1}) as compared
to the dust emission (red contours) mapped in continuum at 1.37\,mm by  \citet{Beuther2018}. Red contours start from a threshold of 27\,mJy\,beam$^{-1}$ increasing by a
factor 2. The equatorial reference frame is rotated clockwise by $44^{\circ}$ to align the (projected) outflow direction with the vertical axis of the plot (linear scale to the
bottom). Large grey spots mark the loci where side spectra were integrated. \textbf{Side panels:} spectra of the local NH$_3$\,(5,\,5) emission (grey histograms) extracted
at the position of the peaks of emission and absorption. In each panel, the blue profile draws the best fit spectrum to the NH$_3$ emission with fitted parameters of excitation
temperature (T) and column density (N) written in the corners. The systemic velocity (V$_{\rm sys}$) of circumstellar gas is marked with a red, vertical, dotted line and the 3D
molecular structure is sketched.}
\label{fig4}
\end{figure*}

The relative brightness of the centimeter ammonia lines provides a sensitive thermometer for the circumstellar gas, as emission at different
excitation energies can be used to find the best description for the gas conditions locally, under the assumption of thermodynamic equilibrium 
\citep[e.g.,][]{Ho1983,Walmsley1983,Mangum2013}. This assumption is satisfied at the high particle densities (in excess of 10$^{7}$\,cm$^{-3}$) expected 
around a massive star across its equatorial plane, at radii of thousand au and less \citep[e.g.,][]{Sanna2021}. In Fig.\,\ref{fig4}, we summarize this analysis
based on the line brightness ratio of the (2,\,2), (4,\,4), and (5,\,5) ammonia para lines. At the frequency of each ammonia transition, we produced integrated
spectra extracted from three pointings placed at the peaks of emission and absorption (grey circles in Fig.\,\ref{fig4}), that are representative of gas seen at the
two opposite sides with respect to the central star and along its direction, respectively (the latter sketched in Fig.\,\ref{fig3}). In the contour map of Fig.\,\ref{fig4},
we also show that the spatial extent of ammonia correlates with the morphology of dust emission (red contours) imaged at comparable resolution by
\citet{Beuther2018}, likely indicating copious evaporation from dust grain mantles where ammonia molecules are synthesized. As an example, side panels in
Fig.\,\ref{fig4} display spectra of the (5,\,5) transition at the higher excitation temperature of 296\,K, whereas blue profiles draw final synthetic spectra obtained 
from the line fitting (best fit parameters in the corners). 

The spectral line fitting delivers twofold information, about the NH$_3$ physical conditions and its kinematics too (Tab.\,\ref{tpar}). Firstly, at the eastern
and western sides of the accretion disk, at an average radius of 380\,au  from the central star, gas is warmed up to comparable temperatures of approximately
170\,K, although the eastern side attains twice as much column (thus material) and linewidth than the western side. This discrepancy cannot be reconciled even
within a 3\,$\sigma$ uncertainty. This finding was already noticed in the past and interpreted as evidence of two distinct hot molecular cores, missing information about
the kinematic pattern of gas at high angular resolution \citep{Brogan2007,Comito2007}. Our observations can now separate the line emission and absorption properly
and probe a continuous physical structure instead. In turn, these results suggest a much higher gas turbulence on one disk side, which, as we will show, has a
counterpart in the smoothed velocity gradient observed to the east of the star than to the west (see Sect.~\ref{toy}). 

Secondly, by looking through the column of NH$_3$ gas directly in front of the young star, we can distinguish two, blended, velocity components, whose peak velocities
are both redshifted with respect to the systemic velocity and whose physical conditions differ. Labeled as component 2, the most redshifted component  (by 5.7\,km\,s$^{-1}$)
is warmer and has a column density almost two times larger than the colder, less redshifted component (by 0.4\,km\,s$^{-1}$), labeled as component 1. This evidence
suggests we are probing two different portions of gas that overlap along the line-of-sight, with the warmer (closer to the star) gas component infalling at higher velocities
in the star direction than the colder component. This scenario of infalling gas can be better understood when we add the kinematic information coming from the first-moment
ammonia map, which, in turn, provides a second indirect proof of infall and is discussed in the following section.

\subsection{Kinematic toy model}\label{toy}

\begin{figure*}
\centering
\includegraphics [angle=0, scale= 0.95]{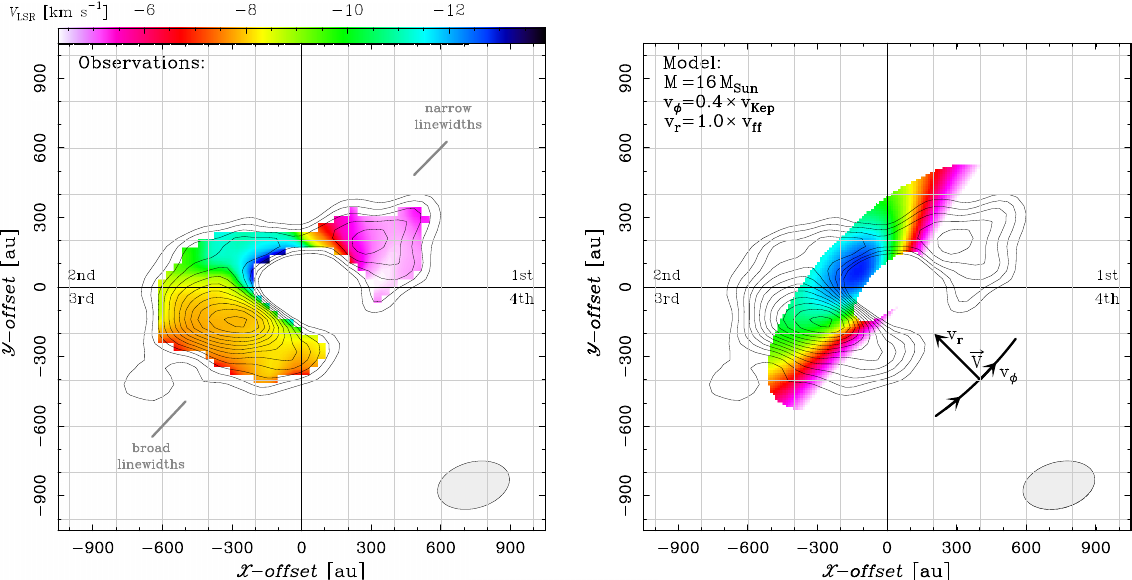}
\caption{Comparison between observations (left panel) and a toy model (right panel) which reproduces the line-of-sight velocity pattern of ammonia. 
\textbf{Left:} Same as the bottom panel of Fig.\,\ref{fig1}, but with axes labeled in astronomical units and offsets counted from the star position. A grid
is drawn at steps of 200\,au for a direct comparison with the right panel and the quadrants are labeled. The third quadrant shows a velocity gradient 
much smoother than in the first quadrant due to ammonia linewidths being two times broader (Fig\,\ref{fig4}). \textbf{Right:} Similar to the left panel,
but with the color map obtained from the toy model of a razor thin disk with the same geometry of circumstellar gas around HW2, and convolved with
the same instrumental beam. Model parameters are written in the top left with the velocity vectors sketched in the fourth quadrant.}
\label{fig5}
\end{figure*}


In order to interpret the kinematic pattern of NH$_3$ gas around HW2, we have built up a toy model presented in the right panel of Fig.\,\ref{fig5}. This model
reproduces the velocity field of an ideal, razor-thin disk that has a radius comparable to the extent of the ammonia emission (700\,au) and where we mask out 
the inner region of absorption (200\,au). The local velocity vector computed from the model is the composition of two free parameters of counterclockwise rotation
and radial motion of gas collapse. The magnitudes of rotation and infall are given as fractions of the local values of Keplerian ($v_{K}$) and free-fall (v$_{ff}$)
velocities, respectively, which only depend on the gravitational field strength centrally. The observer measures  the velocity component projected along the
line-of-sight  (V$_{\rm LSR}$), via  line's Doppler shift, which we analytically derive after taking into account the orientation of the disk plane in space (by setting
the disk inclination, $i$\,$=$\,$64\degr$,  and position angle, P.A.\,$=$\,$134\degr$, estimated from \citealt{Sanna2017}). 
For a direct comparison with observations, we also (i) account for the same systemic velocity, (ii) we smooth the velocity pattern spatially within a kernel equal
to the beam size, and (iii) only plot the range of integrated velocities exploited in the first moment map.

We searched for couples of rotation (v$_{\phi}$) and infall (v$_{\rm r}$) values whose line-of-sight projection reproduces the observed first moment map. 
Their values were  chosen among realistic combinations of Keplerian (V$_{\rm Kep}$) and free-fall (V$_{\rm ff}$) fractions, respectively, whose magnitudes
are functions of the disk radius and the central mass only. The analysis was carried out by considering two possible values for the central mass inside the
absorption region, of 12 and 16\,M$_{\odot}$, which best fit the spectral energy distribution towards Cepheus\,A as measured by \citet{DeBuizer2017}.

In Figs.\,\ref{figA3} and~\ref{figA4}, we show a number of combinations for the percentage of Keplerian and free-fall velocities producing a given kinematic 
pattern along the line-of-sight (V$_{\rm LSR}$). These combinations are explanatory of how the V$_{\rm LSR}$ changes across the ring by increasing (and
decreasing) either one or the other quantity.  There are two characteristic features that constrain the underlying velocity field once the central mass is known,
namely, the position angle of the region of (relatively) bluish emission and the magnitude of the line-of-sight velocities. We explicitly note that, according to the
colored scaled used, regions of reddest emission are those where the line-of-sight velocity is comparable to the systemic velocity, of $\rm -4.5\,km\,s^{-1}$, and
thus the gas motion in the local standard of rest is nearly zero along the line-of-sight. As a rule of thumb, we note that: (1) when decreasing the percentage of
Keplerian velocity, the blueish region of emission shifts clockwise along the ring, while (2) when increasing the percentage of free-fall velocity, the blueish velocities
increase in absolute magnitude. Our observations show that blueish velocities are bracketed inside the second quadrant with an abrupt reddish gradient crossing
over the first quadrant (Fig.\,\ref{fig2}).

This qualitative analysis shows that each different combination of infall and rotation around HW2 produces an unique velocity pattern, that can be compared
with the observations to explain the gas motion around the young star. At first sight, this comparison allows us to exclude, for instance, the possibility of 
fast rotation close to centrifugal equilibrium for radii larger than 200\,au, while it also implies the existence of large infall motions dominating the gas kinematics at
this distance from the star. In particular, on the right panel of Fig.\,\ref{fig5} we list the set of parameters which best match the observed ammonia map (left panel). 
The ammonia pattern is reproduced for gas that is collapsing close to free-fall and rotating at $40\%$ ($\pm10\%$) the Keplerian value at a given radius. 
In fact, by assuming higher fractions of Keplerian rotation, blueish velocities largely encompass the third quadrant as well, at variance with observations, whereas
values lower than $30\%$ the Keplerian rotation underestimate the region of blueish emission (Figs.\,\ref{figA3} and~\ref{figA4}). Our toy model also supports that the
velocity magnitudes can be better reproduced by setting a central mass of 16\,M$_{\odot}$  (Fig.\,\ref{figA4}), as opposed to a smaller value of 12\,M$_{\odot}$ which
fails to (Fig.\,\ref{figA3}). Accordingly, \citet{DeBuizer2017} obtained  lower $\chi^2$ fitting values for a stellar mass of 16\,M$_{\odot}$ than 12\,M$_{\odot}$, by modeling
the spectral energy distribution from Cepheus\,A.

To further support the model consistency, in the left panel of Fig.\,\ref{figA5} we show the full range of velocities predicted to the south-west of the star position, where
the absorption feature suggests strongly redshifted motions. According to the toy model, in this spatial region the local velocity vector should be almost aligned
with the line-of-sight and oriented away from the observer, with velocities higher than $-2.5$\,km\,s$^{-1}$ and below $+5$\,km\,s$^{-1}$ approximately.   
 This expected velocity range corresponds to that covered by the ammonia absorption, in agreement with our observation (Fig.\,\ref{figA1}).

Interestingly, looking at the left panel of Fig.\,\ref{fig5}, the third quadrant shows a much smoother velocity gradient than the first quadrant, as compared to the
model expectations to the right. This finding is directly related to the broad linewidths detected on the eastern side of the accretion disk with respect to the
western narrow linewidths. Because both disk sides have similar temperatures, this difference in linewidth has to be related to a higher degree of gas turbulence
on the eastern side. Moreover, the higher gas column density on the eastern disk side indicates a gas density approximately twice as high, assuming that the disk
structure is mostly uniform all around and that the relative abundance between NH$_3$ and H$_2$ molecules is comparable on both disk sides. We suggest that
the cause behind this difference might be an external injection of mass to the eastern disk side. In this scenario, the accretion disk must be continuously 
replenished of fresh material to sustain the ongoing mass infall towards HW2. Indeed, in the recent literature there is growing evidence for streamlines
of  gas and dust, referred to as streamers, connecting the inner circumstellar regions of young stars with the surrounding envelope 
 \citep[e.g.,][]{Akiyama2019,Pineda2020,Garufi2022}. In this respect, and in support of our hypothesis, we note that a clear streamline of molecular gas  is evident
 in the sub-arcsecond millimeter maps  by \citet[][their Figs.\,2 and~A.1]{Ahmadi2023}.  This emission appears to connect with the eastern side of the 
 accretion disk surrounding HW2.

Finally, we explicitly remind that the purpose of the toy model represented in Fig.\,\ref{fig5} is to reproduce the pattern of velocities observed in ammonia with
 basic assumptions. These assumptions cannot account for the spatial morphology and brightness of ammonia emission whose explanation, in terms of physical and
environmental conditions, goes beyond the scope of the current toy model. On the contrary, this toy model is highly valuable in guiding us to constrain a "more realistic"
scenario which is discussed via independent simulations in the following.

\subsection{Simulations}\label{sim}

\begin{figure*}
\centering
\includegraphics [angle= 0, scale= 1.0]{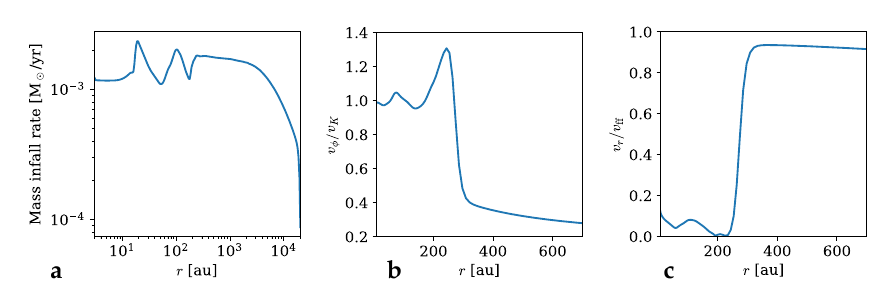}
\caption{Kinematic properties of the accretion disk according to the simulation that resembles our observations the most, taken after 
$t=9.62\,\mathrm{kyr}$ from its beginning. Form left to right, panels show the dependence with radius of the (a) mass infall rate, the (b)
azimuthal velocity as a fraction of the Keplerian value, and the (c) radial velocity as a fraction of the free-fall value.} 
\label{fig6}
\end{figure*}


\begin{figure*}
\centering
\includegraphics [angle= 0, scale= 0.5]{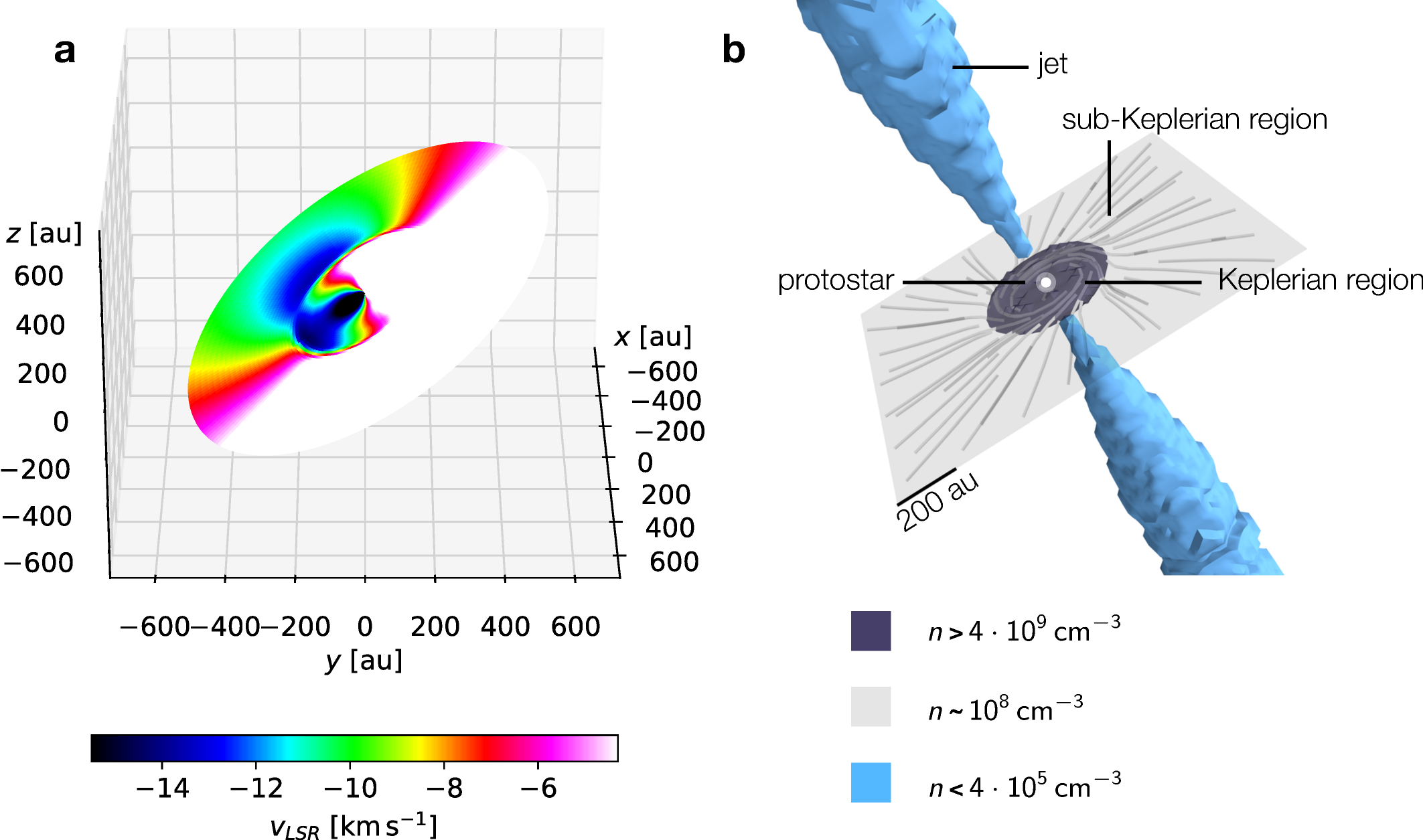}
\caption{Three-dimensional view of the accretion disk kinematics as obtained from the selected simulation, taken after $t=9.62\,\mathrm{kyr}$ from
its beginning. To the left, panel\,(a) shows the theoretical pattern of line-of-sight velocities within a similar range as Fig.\,\ref{fig5} for a direct comparison,
including also the inner disk regions where the rotation curve becomes Keplerian  (masked by absorption in our observations). To the right, panel\,(b)
shows the entire disk and jet structures produced by the selected simulation, where reference levels of hydrogen volume density are listed for the
accretion disk (Keplerian and sub-Keplerian regions separately) and the jet.} 
\label{fig7}
\end{figure*}


Notably, we find that the kinematic pattern observed around HW2 is indeed predicted by recent, ab initio, simulations for the formation of stars as massive
as HW2. In order to provide a theoretical ground validating our interpretation independently, we searched a recently published catalog of numerical simulations
\citep{Oliva2023a,Oliva2023b}. These simulations start from the gravitational collapse of a cloud core threaded by an uniform magnetic field initially, whose
outcome will be a massive (proto)star surrounded by an accretion disk and a magnetically-driven outflow eventually. Gas and dust were modeled along with
the stellar evolution \citep{Hosokawa2009} by accounting for non-ideal magneto-hydrodynamic effects \citep{Mignone2007}, Ohmic dissipation 
\citep{Machida2007}, self-gravity \citep{Kuiper2010}, and radiation transport approximated via gray flux-limited diffusion \citep{Kuiper2020}. The spatial domain
was sampled through an axis-symmetric grid in spherical coordinates, with 224 and 40 cells in the radial and polar directions respectively, and increasing the
radial dimension logarithmically with distance.

Simulations in the catalog explore several possible pre-collapse conditions for massive star formation, varying the initial mass of the cloud, density and angular
momentum distributions,  and the initial magnetic field strength. In order to identify a set of simulations approximating our observations more closely, we constrained 
candidates to have an accreted stellar mass similar to HW2, with a Keplerian disk size equal or smaller than 200\,au, as well as to drive a highly collimated jet.
The best match was found for an initial cloud core of $100\,\mathrm{M_\odot}$ enclosed within $0.1\,\mathrm{pc}$, whose density distribution is a power law
with radius as, $\rho \propto r^{-2}$. The angular momentum is calculated assuming initial solid body rotation with a rotational-to-gravitational-energy ratio 
of $4\%$. The initial magnetic field strength is set to $68\,\mu\mathrm{G}$ corresponding to a mass-to-flux ratio of 20 times the critical value (below which magnetic
fields prevent collapse).

The steep density profile triggers the cloud to rapidly collapse when the system is left to evolve freely, while the angular momentum transport towards the center
slowly builds up an accretion disk. After $9.62\,\mathrm{kyr}$ from the start, the young star has already reached a mass of $16\,\mathrm{M_\odot}$ consistent 
with that of HW2. Despite the high  stellar mass, the system is still young (namely, without photo-ionization) because of the strong accretion rates reached at
this early stage that, in turn, depend on the initial conditions of the collapse. The current mass of the accretion disk amounts to 2.2\,M$_{\odot}$ inside a radius
of 700\,au, of which 1.6\,M$_{\odot}$ are enclosed within a Keplerian radius of 240\,au only, and indicates a disk-to-star mass ratio of 10--14\%. This conclusion
is in good agreement with recent estimates of the inner gas mass surrounding HW2, between 0.8--2.6\,M$_{\odot}$, as obtained through interferometric
observations of millimeter dust emission \citep{Beuther2018,Gieser2021,Ahmadi2023}.

In Figs.\,\ref{fig6} and~\ref{fig7}, we plot a number of results from the selected simulation for comparison.  Beginning with Fig.\,\ref{fig6}, the left panel\,(a)
shows the theoretical mass infall rate calculated along the equatorial plane as a function of the radial distance from the star, whose values are in excellent agreement
with our measurement of  $\rm 2.0\cdot10^{-3}\,M_{\odot}\,yr^{-1}$ (see below). In the middle panel\,(b) of the same figure, we show the percentage of Keplerianity 
with radius. The flow in the innermost $\sim$250\,au of the disk is nearly Keplerian and decreases to 30-40\% the Keplerian value for radii between 400 and
700\,au, in good agreement with our findings. In the right panel\,(c), we plot the radial velocity of gas as a function of the disk radius. At a decrease of Keplerianity
corresponds a steep increase of the radial component of gas collapse which approaches theoretical free-fall at 300\,au already, in good agreement with our observations.
Finally, in Fig.\,\ref{fig7} we produced a three-dimensional synthetic image of the velocity along the line-of-sight ($\rm v_{LSR}$), tracing material surrounding the
young star within 700\,au and using a similar color scale as in Fig.\,\ref{fig5}. In the spatial range traced by ammonia emission (and absorption), the simulations,
toy model and observations are all consistent with each other.  In particular, the left panel of Fig.\,\ref{fig7} shows the sharp transition in the disk pattern
where the gas kinematics changes from a sub-Keplerian regime to a Keplerian one within about 200\,au from the star. For completeness, in the right panel of
Fig.\,\ref{figA5} we also show the full kinematic pattern of gas around the young star, tracing the higher velocities (in absolute value) that are associated with the inner 
Keplerian rotation.

\section{Mass infall rate in the inner disk regions}\label{Mdot}

Having information about the magnitude of infall velocity at a given disk radius, together with the physical conditions of local gas, we are in a position
to calculate the amount of gas mass crossing the disk per unit time towards the central star, as follows. We firstly consider a flow of mass M and average 
density $\rho$, moving through a spherical shell with radius R and thickness dR over a time intervall dt. The total mass crossing the shell per unit time is,  
$\rm \dot{M} = 4\,\pi\,R^2 \cdot (dR/dt) \cdot \rho$. Now, we consider an inward motion with $\rm v_r = dR/dt$ that only flows through a small solid angle
$\Omega$, such as the opening angle of a flared disk. Then, we can rewrite $\rm \dot{M}$ in convenient units:
{\small
\begin{equation}\label{Min}
\left[\rm \frac{\dot{M}_{in}}{M_{\odot}\,yr^{-1}}\right]=
1.5\cdot10^{-4} \left[\rm \frac{R}{100\,au}\right]^2 \cdot \left[\rm \frac{v_r}{10\,km\,s^{-1}}\right] \cdot \left[\rm \frac{n_{H_2}}{10^8\,cm^{-3}}\right] \cdot \left[\frac{\Omega}{4\pi}\right]
\end{equation}
}
where $\rm n_{H_2}$ refers to the volume density of molecular hydrogen, that is the most abundant molecule by orders of magnitude.

In equation\,\ref{Min},  each variable can be quantified based on the results of our observations. We solve eq.\,\ref{Min} for a distance $\rm R_{x}$\,$=$\,380\,au at
the position of our western measurement of NH$_3$ physical conditions. At this distance from the star, the theoretical free-fall velocity amounts to 8.6\,km\,s$^{-1}$.
In order to estimate the volume density of molecular hydrogen at $\rm R_{x}$, we refer to the sketch in Fig.\,\ref{figA6} and reason as follows. As a common assumption,
we consider a flared disk where the ratio between the disk height ($\rm H_{i}$) at varying radii ($\rm R_{i}$) is constant, $\frac{\rm H_i}{\rm R_i}=\tan{\alpha}=K$.  The 
disk height along the line-of-sight through $\rm R_{x}$ is,  $\rm H_{x}$\,$=$\,$\rm \sqrt{\rm R_x^2+x^2}  \cdot K$, and depends on the projection x along the disk plane. 
Alternatively, we can  write the ratio between $\rm H_{x}$ and x as:

{\small
\begin{equation}\label{theta}
\frac{\rm H_x}{\rm x}=\tan{\theta}=\sqrt{1+ \frac{\rm R_x^2}{\rm x^2}}  \cdot \rm K
\end{equation}
}

Notably, the right hand side of this equation is always greater than $K$, meaning that (1) the angle $\theta$ is greater than $\alpha$, which is the flaring angle,
and  that (2) $\theta$ increases as the projection x decreases. Also, x has an upper limit set by the geometry of the ammonia disk ($\rm R_{NH_3}$\,$=$\,700\,au)
and the position of our measurement ($\rm R_{x}$\,$=$\,380\,au), so that, $\rm x_{max}$\,$=$\,$\rm \sqrt{\rm R_{NH_3}^2-R_{x}^2}$\,$=$\,587\,au and
$\rm \tan{\theta} = 1.19 \cdot K$.  Thus, we can calculate an upper limit to the disk path crossed by the line-of-sight through the $\rm R_{x}$ position, of:

{\small
\begin{equation}\label{path}
\rm 2 \times p_{x} < 2 \cdot \frac{x_{max}}{\cos{26\degr}} = 1306\,au
\end{equation}
}
where the factor 2 accounts for both positive and negative heights with respect to the disk mid-plane. In conclusion, by dividing the (lower) western column
density of $3.98\cdot10^{16}$\,cm$^{-2}$ for this path, and accounting for a NH$_3$-to-H$_2$ abundance ratio of $3\cdot10^{-9}$, we derive a lower limit
to the local H$_2$ volume density of $6.8\cdot10^{8}$\,cm$^{-3}$. The abundance ratio is a conservative upper limit inferred by comparing the higher value
of NH$_3$ column, of $7.08\cdot10^{16}$\,cm$^{-2}$, with the peak column density of molecular hydrogen, of $2.1\cdot10^{25}$\,cm$^{-2}$, as measured
at comparable resolution through millimeter continuum emission \citep[][their Table\,A.1]{Beuther2018}. 

The column of warm ammonia gas that we detect in absorption along the line-of-sight comprises gas coming from an outer disk layer above the mid-plane
by $26\degr$ (Fig.\,\ref{fig3} and~\ref{figA6}). Its column density is more rarefied by an order of magnitude than that of the emission measured by looking
through the disk plane directly (at R$_x$\,$=$\,\,380\,au in Fig.\,\ref{figA6}), although the path intercepted by the line-of-sight to the center is of the same order 
of magnitude as p$_{max}$. This evidence suggests that the disk should be flared by an opening angle $\Omega$\,$\leq$\,$2 \times 26\degr \approx 60\degr$.

Finally, plugging in these numbers in eq.\,\ref{Min}, we derive that at an average distance of 380\,au from the star, the inward stream of gas, flowing at a radial
velocity of $\rm 8.6\,km\,s^{-1}$, induces a mass infall rate of $\rm 2.0\cdot10^{-3}\,M_{\odot}\,yr^{-1}$. This result only decreases by a factor of 1.5 when replacing
the theoretical free-fall velocity with the peak value of the most redshifted component measured along the line-of-sight (Tab.\,\ref{tpar}), of 5.7\,km\,s$^{-1}$ (being
a line-of-sight average).

\section{Conclusions}\label{concl}

We report on spectroscopic, B-configuration, Jansky Very Large Array (VLA) observations of hot ammonia emission near 1\,cm towards Cepheus\,A HW2. 
We target the metastable inversion transitions up to the (5,\,5) line and map a ring of ammonia gas between radii of about 700\,au and down to 200\,au from
HW2, as part of its accretion disk. Ammonia is detected both in emission and absorption, the latter seen in foreground against the bright radio continuum of
the radio thermal jet driven by HW2.

We resolve the gas kinematics across the ring which we interpret via both a toy model and a direct comparison with independent ab-initio simulations. 
This analysis shows that the velocity field traced by ammonia is a composition of slow rotation and prominent infall motions that still overtake centrifugal equilibrium
at disk radii of a few 100\,au only. At this distance from the star, gas slowly orbits at a few 10\% the Keplerian value around a central mass of 16\,M$_{\odot}$, consistent
with the spectral energy distribution from the region. In particular, our simulations predict a similar scenario as observed around HW2 when the system is still very young,
of the order of 10\,kyr, and a Keplerian regime is only reached in the innermost 200\,au accordingly.

We measure ammonia physical conditions across the ring by fitting together the (2,\,2), (4,\,4), and (5,\,5) ammonia para lines and inferring gas temperatures  
above 170\,K for disk radii smaller than 380\,au. This analysis also suggests a higher degree of gas turbulence on one disk side than to the other, which might 
be related  to a streamer of dense gas detected in previous millimeter observations of Cepheus\,A. In particular, we analytically derive the mass infall rate of
gas streaming through the disk at a radius of 380\,au, whose high value, of $2\times10^{-3}$\,M$_{\odot}$\,yr$^{-1}$, agrees with that predicted by our 
independent simulations.

Finally, we remind that two independent detections of luminosity bursts have been reported from as many young and massive stars \citep{Caratti2017,Hunter2017}.
These observations are proof of  episodic events of stellar accretion, where a circumstellar disk can suddenly convey onto the star a mass as large as few Jupiter
masses  ($\approx3\times10^{-3}$\,M$_{\odot}$). If these findings imply the existence of an accretion disk indirectly, they also require mass infall rates of the same
order of the accreted mass. Indeed, we prove observationally that similar infall rates can be efficiently sustained by an accretion disk, such as the one we observe  
around the nearby HW2. In doing so, we settle a dispute that has long questioned the existence of an accretion disk around one of the closest massive
forming stars to the Sun, which, because of HW2 preferred position, had cast doubts on the accretion mechanism itself.

\begin{acknowledgements}

Comments from an anonymous referee, which helped improving our paper, are gratefully acknowledged.
This paper makes use of data obtained with the Karl G. Jansky Very Large Array, operated by the National Radio Astronomy Observatory
(NRAO). The NRAO is a facility of the National Science Foundation operated under cooperative agreement by Associated Universities, Inc.
This paper makes use of the Grenoble Image and Line Data Analysis Software (GILDAS) -- https://www.iram.fr/IRAMFR/GILDAS.
A.S. acknowledges financial support from the INAF-Minigrant 2022  ``Feasibility project for the ngVLA: a Science Case with Masers'' (ngVLAmas,
P.I.: A. Sanna). 
A.O. acknowledges financial support from the State Secretariat for Education, Research and Innovation of the Swiss Confederation via their
program ``Swiss Government Excellence Scholarships for Foreign Students 2023/2024", the Office of International Affairs and External
Cooperation of the University of Costa Rica, and the European Research Council (ERC) under the European Union's Horizon 2020 research
and innovation program (grant agreement No. 833925, project STAREX). 
G.S acknowledges the project PRIN-MUR 2020 MUR BEYOND-2p  ``Astrochemistry beyond the second period elements'' (Prot. 2020AFB3FX) 
and the INAF-Minigrant 2023 ``TRacing the chemIcal hEritage of our originS: from proTostars to planEts'' (TRIESTE, PI: G. Sabatini). 
J.M.T. acknowledges partial support from the PID2023-146675NB grant funded by MCIN/AEI/10.13039/501100011033, and by the program Unidad
de Excelencia Mar\'{\i}a de Maeztu CEX2020-001058-M. 
A.C.G. acknowledges support from PRIN-MUR 2022 20228JPA3A ``The path to star and planet formation in the JWST era'' (PATH) funded by
NextGeneration EU, by INAF-GoG 2022 ``NIR-dark Accretion Outbursts in Massive Young stellar objects'' (NAOMY), and by Large Grant INAF
2022 ``YSOs Outflows, Disks and Accretion: towards a global framework for the evolution of planet forming systems'' (YODA).

%

\end{acknowledgements}


\bibliographystyle{aa}
\bibliography{asanna0225}



\begin{appendix}

\section{Additional material}\label{Apx1}


\begin{table*}
\caption{Summary of VLA settings for experiment 19A-248 observed during Spring and early Summer 2019.}\label{tobs}
\centering
\begin{tabular}{ c c c c c c c c c c}

\hline \hline
    Array        & Band  & R.A.\,(J2000)  &        Dec.\,(J2000)        &   V$_{\rm LSR}$   &     $\Delta$\,v    &  Calibrators  &   HPBW    &  1\,$\sigma$  \\
                    &          &     (h\,m\,s)     &  ($^{\circ}$\,$'$\,$''$) &   (km\,s$^{-1}$)   &  (km\,s$^{-1}$)  &                   &    ($''$)   &      (mJy beam$^{-1}$)        \\
    (1)           &   (2)   &        (3)          &              (4)                   &        (5)                &           (6)           &       (7)         &     (8)     &        (9)          \\
\hline
 & & & & & & 3C286 & &  \\
B & K & 22:56:17.9089 & $+$62:01:49.527 &  $-5.0$  & 0.5 & J2148$+$6107 & 0.265 & 1.0 \\
\hline
\end{tabular}
\tablefoot{Columns\,1 and\,2: interferometer configuration and frequency band. Columns\,3 and\,4: equatorial coordinates of target phase center. Columns\,5
and\,6: source radial velocity and maximum channel resolution. Column\,7:  absolute flux (top) and phase (bottom) calibrators observed. Columns\,8 and\,9: nominal
beam size at a representative frequency of 24.14\,GHz, assuming Briggs's robust weighting ($\Re$\,$=$\,0.5), and rms noise required per channel unit.}
\end{table*}



\begin{table*}
\caption{Observed Ammonia lines\label{tlines}}
\centering
\begin{tabular}{c c r}

\hline \hline
\multicolumn{1}{c}{$\nu$}  &  \multicolumn{1}{c}{Line} & \multicolumn{1}{c}{E$_{\rm up}$/$k$}  \\
\multicolumn{1}{c}{(MHz)}  &                                         & \multicolumn{1}{c}{(K)}                         \\ 
\hline

 & & \\

23694.496 & NH$_3$\,(1,\,1), v\,$=$\,0  &  23.8  \\
23722.633 & NH$_3$\,(2,\,2), v\,$=$\,0  &  65.0  \\
23870.129 & NH$_3$\,(3,\,3), v\,$=$\,0  & 124.1  \\
24139.416 & NH$_3$\,(4,\,4), v\,$=$\,0  & 201.1  \\
24532.989 & NH$_3$\,(5,\,5), v\,$=$\,0  & 295.9  \\ 

\hline
\end{tabular}

\tablefoot{Frequencies and upper energy levels for each molecular transition are obtained from the JPL catalogue \citep{Pearson2010}.}\\

\end{table*}


\begin{table*}
\caption{Physical parameters obtained by fitting the NH$_3$ spectra between 23-25\,GHz.}\label{tpar}
\centering
\begin{tabular}{ l c l r l c c l }

\hline \hline
\multicolumn{1}{c}{Label}    & \multicolumn{1}{c}{R$_{\rm p}$}    &     \multicolumn{1}{c}{size}      & \multicolumn{1}{c}{V$_{\rm LSR}$}  &
\multicolumn{1}{c}{FWHM}             & \multicolumn{1}{c}{T$_{\rm ex}$}  &    \multicolumn{1}{c}{N$_{\rm NH_3}$} & \multicolumn{1}{c}{$\tau_{55}$\tablefootmark{c}}  \\
                                            &         \multicolumn{1}{c}{(au)}         &    \multicolumn{1}{c}{($''$)}      &  \multicolumn{1}{c}{(km\,s$^{-1}$)}   & 
\multicolumn{1}{c}{(km\,s$^{-1}$)} &\multicolumn{1}{c}{(K)} &  \multicolumn{1}{c}{(10$^{16}$\,cm$^{-2}$)} &  \\
\multicolumn{1}{c}{(1)} & \multicolumn{1}{c}{(2)} & \multicolumn{1}{c}{(3)}  & \multicolumn{1}{c}{(4)} & \multicolumn{1}{c}{(5)} &
\multicolumn{1}{c}{(6)}  &\multicolumn{1}{c}{(7)} & \multicolumn{1}{c}{(8)}  \\
\hline
 & & & & & & &  \\
Eastern:    & +380  &  0.452\tablefootmark{a} & --8.1$\pm$0.1 & 7.5$\pm$0.3 & $183^{+11}_{-10}$ & $7.08^{+0.25}_{-0.31}$ & 0.2  \\
 & & & & & & &  \\
Western:   & --380  &   0.452\tablefootmark{a} & --4.3$\pm$0.1 & 3.9$\pm$0.2 & $165^{+11}_{-12}$ & $3.98^{+0.24}_{-0.17}$ & 0.2  \\
 & & & & & & &  \\
Central: \tablefootmark{b}     &     0    &  0.452 & --4.1$\pm$0.2 & 4.1$\pm$0.8 & $176^{+15}_{-13}$ &  $0.21^{+0.03}_{-0.02}$ & 0.01 \\
                                              &     0    &  0.452 & +1.2$\pm$0.2 & 5.0$\pm$0.8 & $216^{+16}_{-16}$ &  $0.35^{+0.05}_{-0.04}$ & 0.01 \\

\hline
\end{tabular}
\tablefoot{Columns\,1 to~3:  pointing position with reference to Fig.\,\ref{fig3}, sky projected distance from the star, and extent of integrated spectra,
respectively. Columns\,4 and~5: combined LSR velocity and FWHM of the NH$_3$ para lines with uncertainties, the former to be compared with a
systemic velocity of $-4.5$\,km\,s$^{-1}$. Columns\,6 and~7: excitation (and kinetic) temperature and column density of NH$_3$ gas modeled with 
MCWeeds (with fitted uncertainties). Column\,8: opacity of the NH$_3$\,(5,\,5) line. \tablefoottext{a}{Extended emission.}\tablefoottext{b}{Composition
of two velocity components, one per line.}\tablefoottext{c}{We explicitly note that all fitted components are optically thin.}}
\end{table*}

\clearpage

\begin{figure}
\centering
\includegraphics [angle= 0, scale= 0.85]{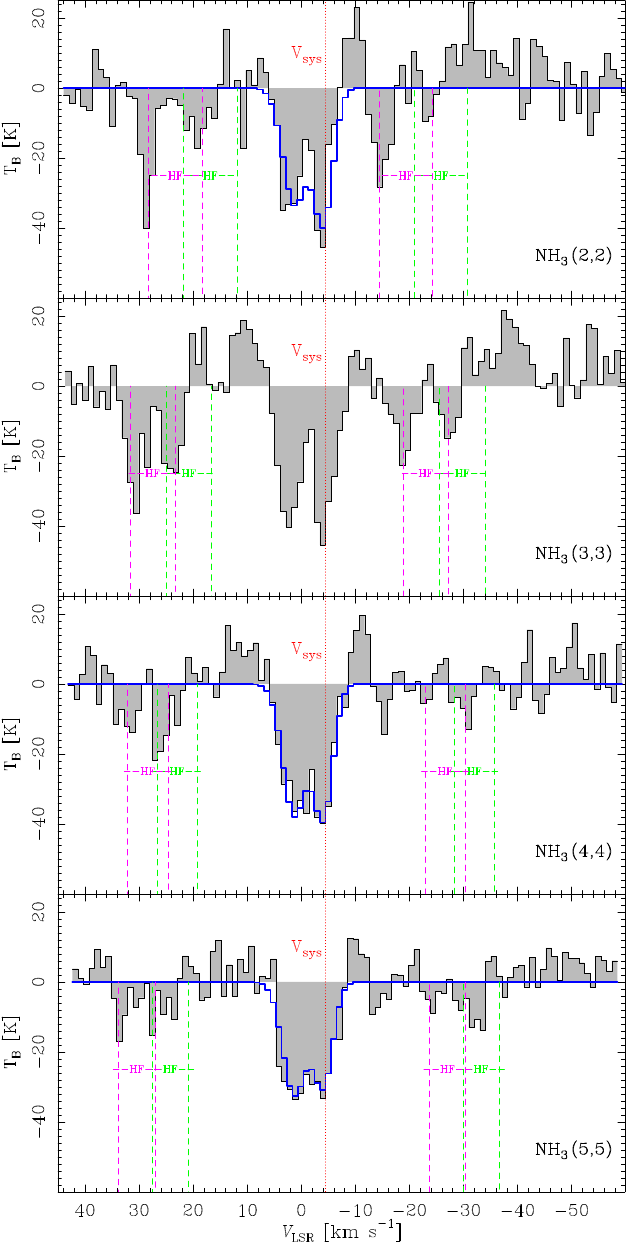}
\caption{Integrated spectra of the NH$_3$ line emission (gray) evaluated over the central absorption region. From top to bottom are displayed spectra of the 
inversion lines from (2, 2) to (5, 5) as labeled on the bottom right. The absorption feature near the systemic velocity of the region, indicated with a dotted red line
(V$_{\rm sys}$), is the combination of two, distinct, velocity components. Theoretical positions of the satellite hyperfine lines (HF) are approximately indicated for
both components in green and magenta separately, whereas the blue profile draws the best fit spectrum to the ammonia para lines.}
\label{figA1}
\end{figure}



\begin{figure}
\centering
\includegraphics [angle= 0, scale= 0.85]{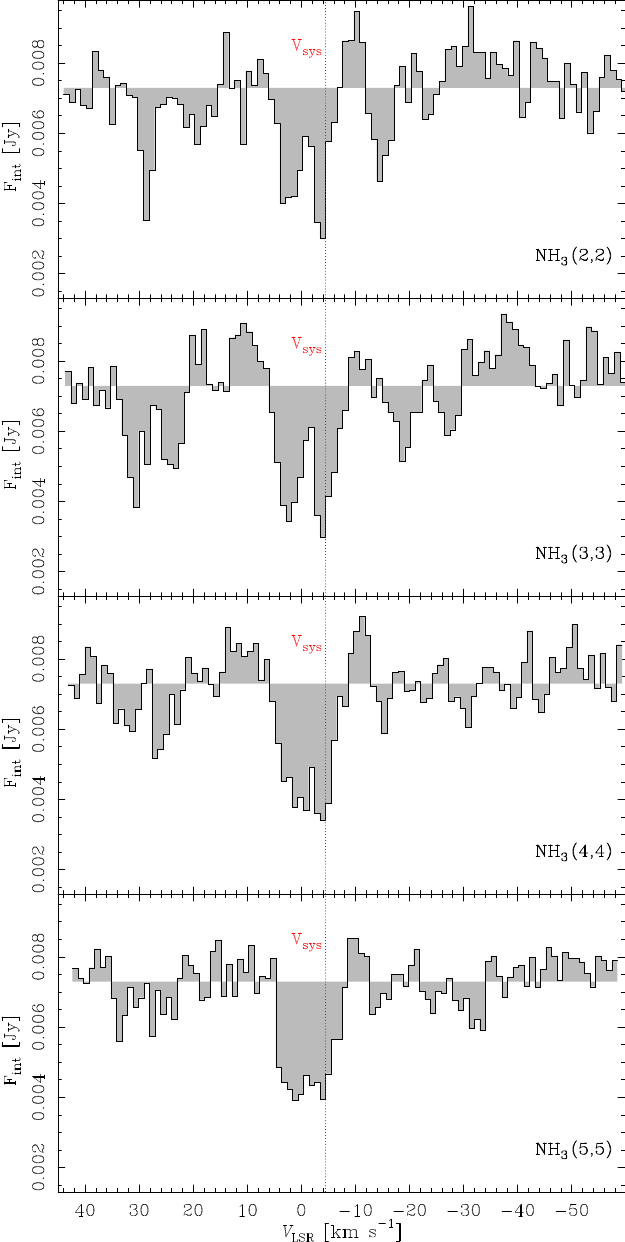}
\caption{Similar to Fig.\,\ref{figA1} but with the continuum level included for comparison.} 
\label{figA2}
\end{figure}



\begin{figure*}
\centering
\includegraphics [angle= 0, scale= 0.82]{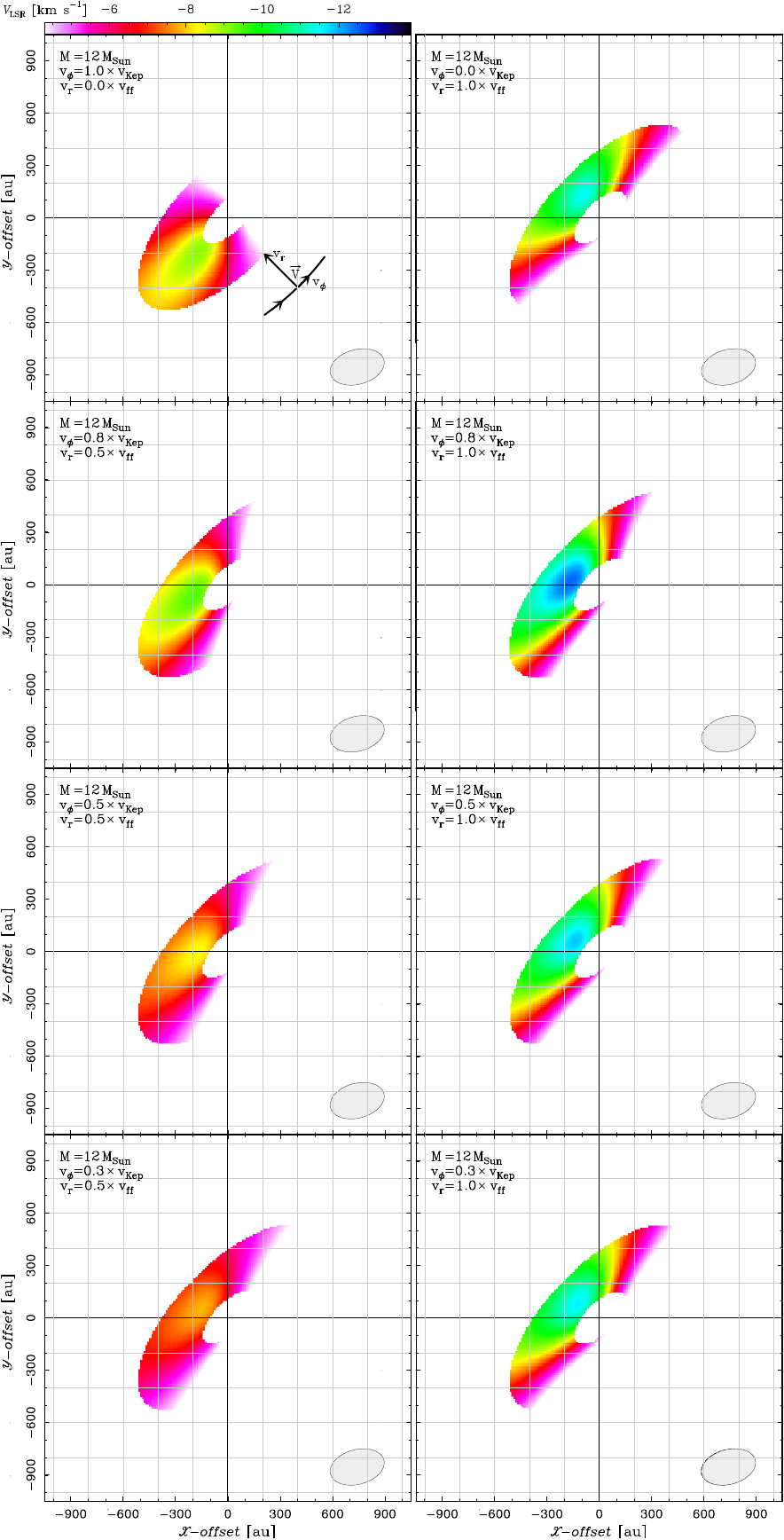}
\caption{Different fractions of Keplerian (v$_{\rm Kep}$) and free-fall (v$_{\rm ff}$)  velocities which can affect the appearance of the kinematic pattern observed
in ammonia, assuming a central mass of 12\,M$_{\odot}$. Same symbols as in Fig.\,\ref{fig5}.} 
\label{figA3}
\end{figure*}



\begin{figure*}
\centering
\includegraphics [angle= 0, scale= 0.82]{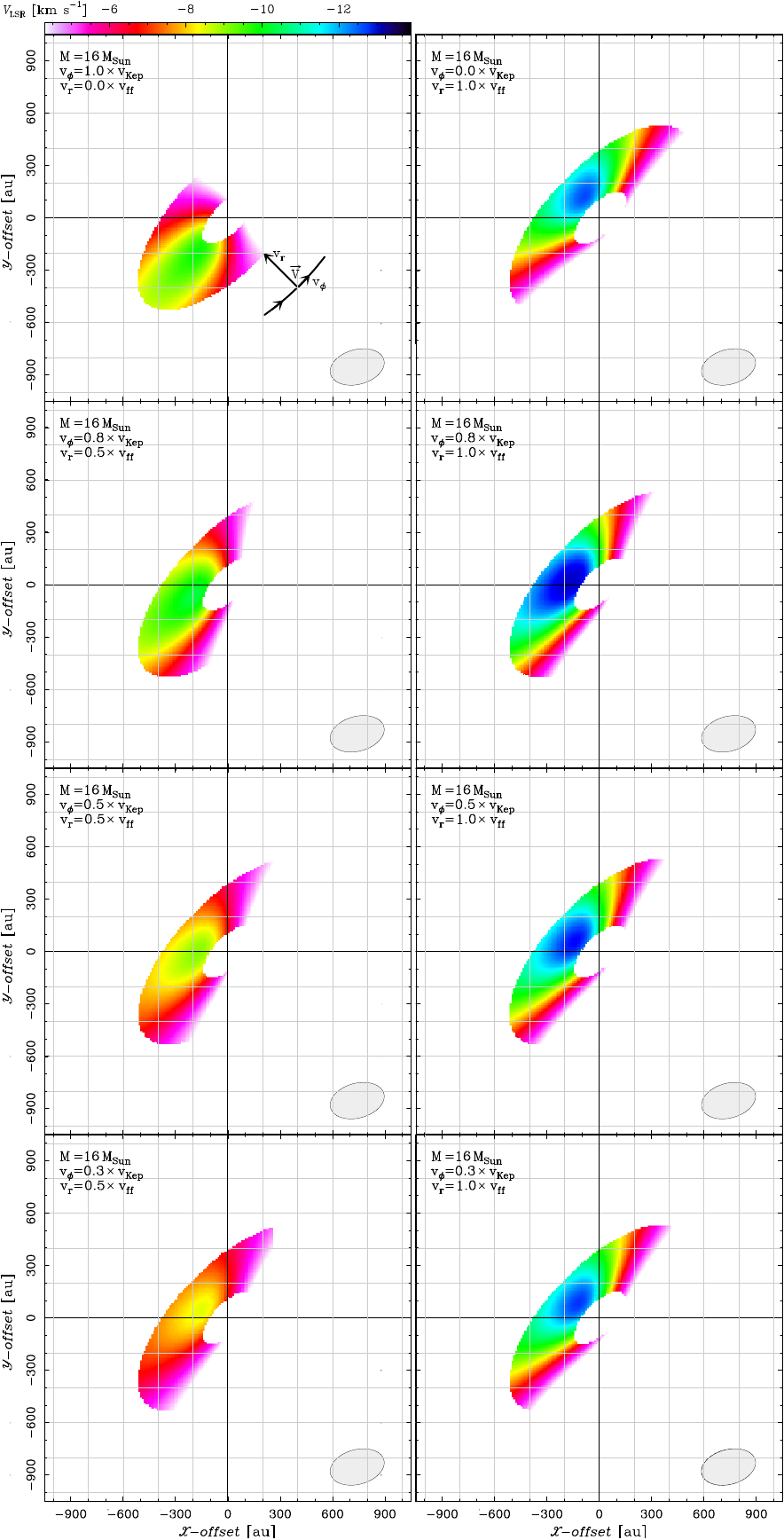}
\caption{Similar to  Fig.\,\ref{figA3}, but for a central mass of 16\,M$_{\odot}$.} 
\label{figA4}
\end{figure*}



\begin{figure*}
\centering
\includegraphics [angle= 0, scale= 1.0]{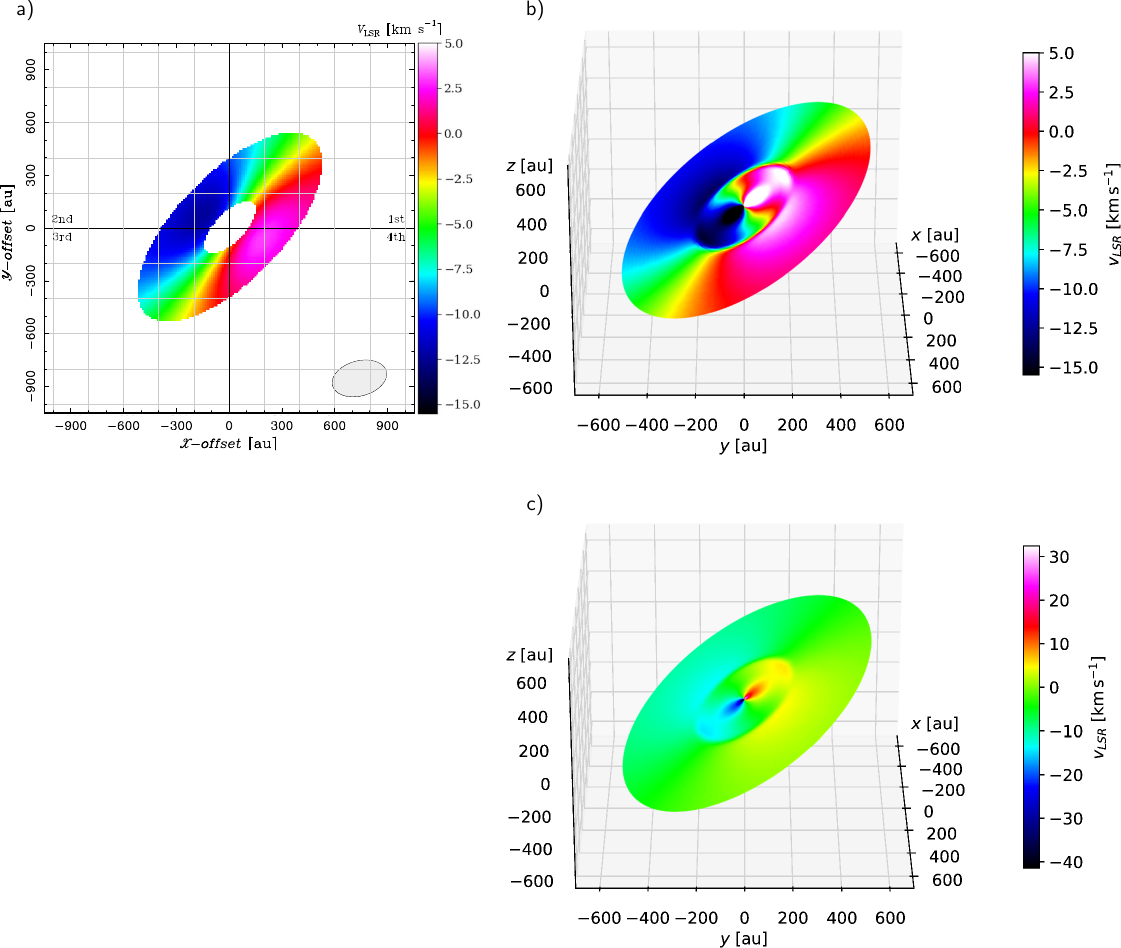}
\caption{Full velocity field of circum-stellar gas around HW2, as predicted by the toy model (panel a) and the simulations (panels b and c). 
These plots are produced under the same conditions as in the right panel of Fig.\,\ref{fig5} (for the toy model) and Fig.\,\ref{fig7}a (for the simulations), but 
show the full range of gas velocities expected theoretically. In particular, in panel a), velocities colored from orange to magenta cover the range of absorption
observed in the ammonia  spectra. Panel b) shows the simulated velocities plotted with the same color scale of the toy model to the left, for comparison,
and we explicitly note that the toy model and simulations predict consistent velocities with each other (cf. Fig.\ref{fig5} and \ref{fig7}a).  Panel c) shows
the simulated velocities with a larger velocity range, to include the innermost and fastest (Keplerian) velocities not covered by the toy model.}
\label{figA5}
\end{figure*}


\begin{figure*}
\centering
\includegraphics [angle= 0, scale= 0.2]{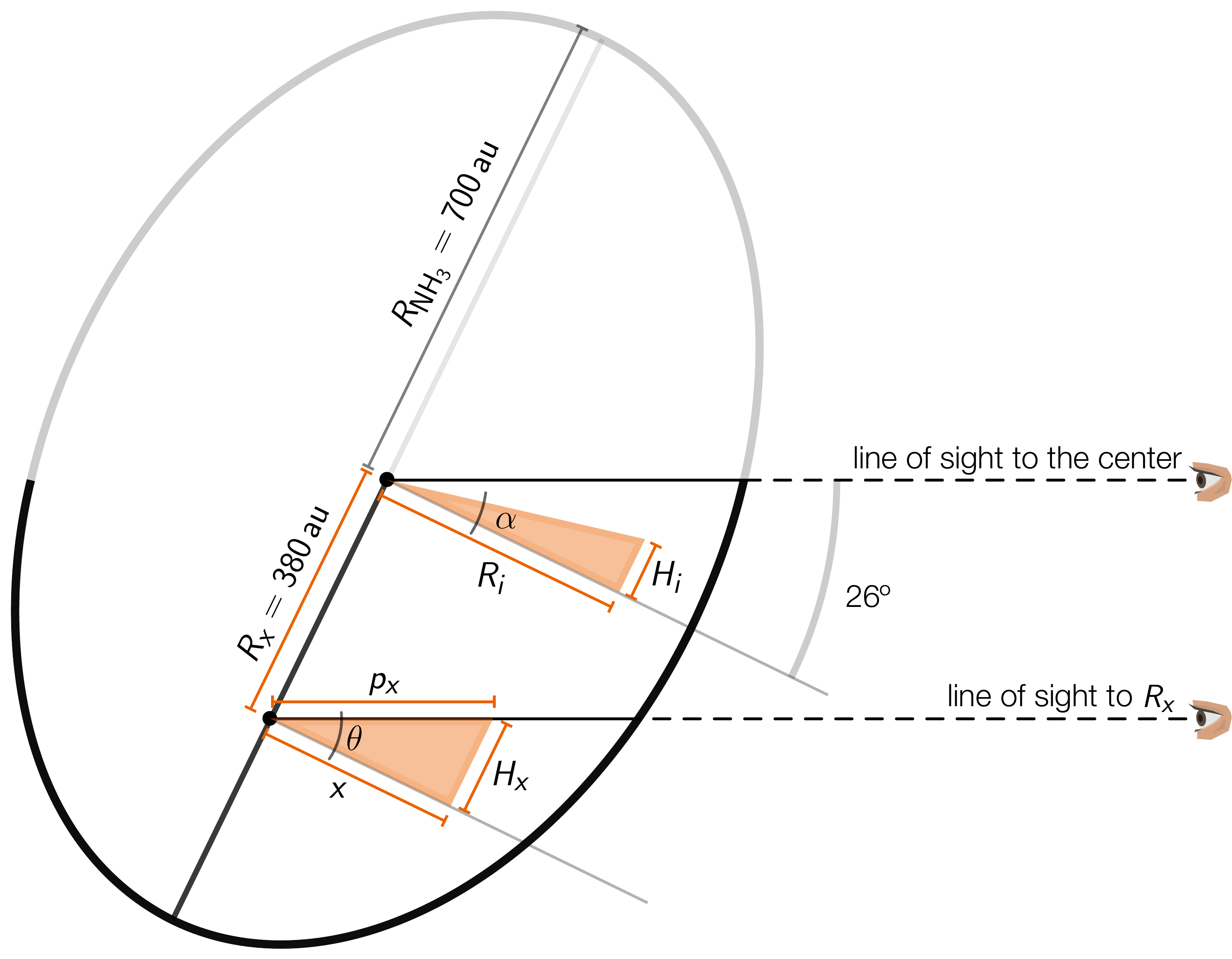}
\caption{Sketch for calculating the maximum path of the NH$_3$ column density measured by an observer who is looking through the 
disk at different positions. For instance, when an observer to the right looks at the position $\rm R_x$ on the eastern disk side, its line-of-sight 
crosses a path which is $\rm 2 \times p_x$ long, where the angle $\theta$ is always greater than the flaring angle $\alpha$.} 
\label{figA6}
\end{figure*}


\clearpage

\section{Radio continuum}\label{Apx2}

\begin{figure*}
\centering
\includegraphics [angle= 0, scale= 0.85]{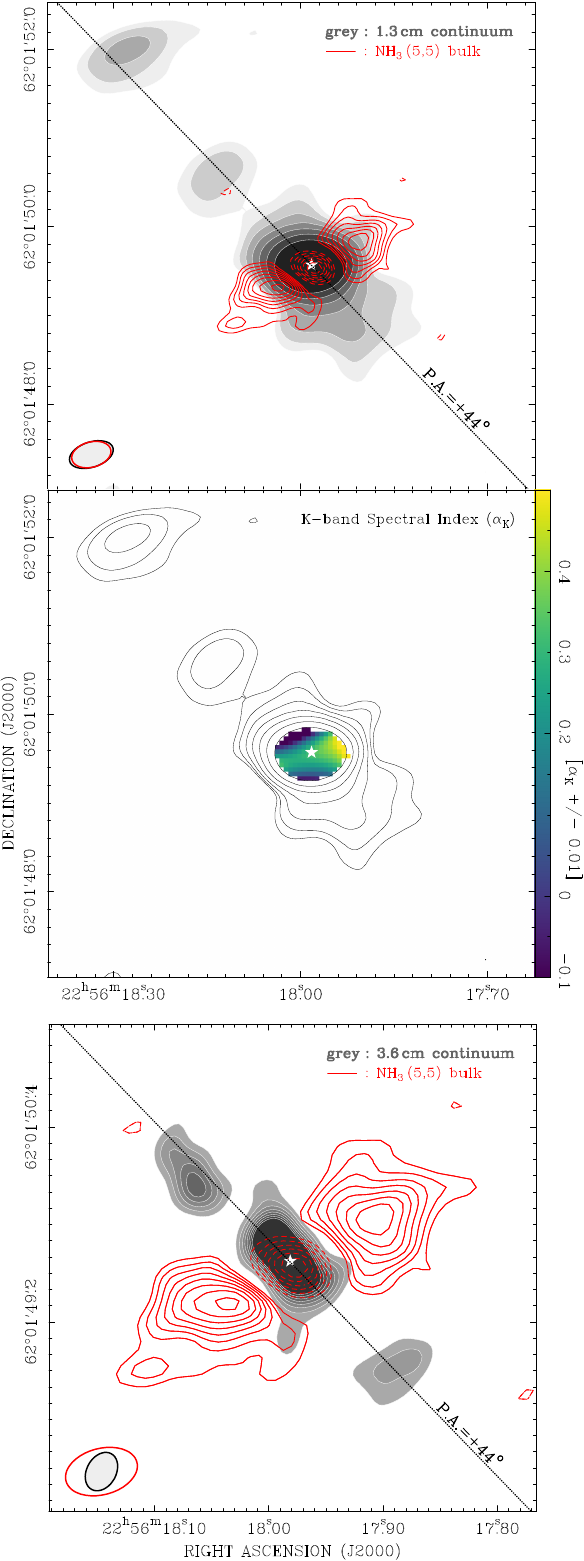}
\caption{Thermal radio jet emission (grey shades) driven by HW2  (white star) compared with the spatial distribution of the NH$_3$ emission (red contours). 
\textbf{Top:} radio continuum emission at 1.3\,cm (K band), obtained from the same NH$_3$ dataset, overplotted on the (5,\,5)
inversion line emission near the systemic velocity. Radio continuum levels start at 7\,$\sigma$ increasing by factors of 2 (1\,$\sigma$ of 5\,$\mu$Jy\,beam$^{-1}$).
The dotted black line traces the mean jet axis at a position angle of 44\,$\degr$ for comparison. Synthesized beams on the bottom left.
\textbf{Middle:} In-band spectral index map (colors) measured from the K-band radio continuum (black contours, same as upper panel). Values according to the color 
wedge to the right.
\textbf{Bottom:} Similar to the upper panel but for the radio continuum emission at 3.6\,cm (X band, see Appendix\,\ref{Apx2}) and a field of view two times smaller. 
Radio continuum levels start at 5\,$\sigma$ increasing at steps of 3\,$\sigma$ (1\,$\sigma$ of 36\,$\mu$Jy\,beam$^{-1}$).}
\label{figB1}
\end{figure*}

In Figure\,\ref{figB1}, we present a summary of the radio continuum imaging. The top panel  shows  the appearance of the continuum map at 1.3\,cm, as obtained from
the line free channels of the current ammonia dataset. The middle panel shows the spectral index map of the continuum at 1.3\,cm in the region of ammonia absorption,
where the nominal uncertainty is better than 0.01 thanks to the high signal-to-noise ratio. The lower panel shows the appearance of the continuum map at 3.6\,cm from
archival data  \citep{Curiel2002}, whose angular resolution is two times better than that of the upper panel and whose absolute position has been registered for the secular
motion of the star forming region \citep[as derived from][]{Moscadelli2009}. The spectral index map, generated following \citet{Sanna2018}, shows values between $-0.1$ 
and $0.5$ as expected for a radio thermal jet. This result confirms that the bright continuum background, against which ammonia lines are observed in absorption,
originates from the compact ejected gas driven by HW2 \citep{Carrasco2021}.

\end{appendix}

\end{document}